# Analyzing the World-Wide Impact of Public Health Interventions on the Transmission Dynamics of COVID-19

George Mohler, Indiana University - Purdue University Indianapolis, gmohler@iupui.edu
Frederic Schoenberg, University of California Los Angeles, frederic@g.ucla.edu
Martin B. Short, Georgia Institute of Technology, mbshort@math.gatech.edu
Daniel Sledge , University of Texas Arlington, dsledge@uta.edu

The Chinese government began implementing extensive public health measures aimed at confronting COVID-19 during January 2020. These measures included contact tracing, disease surveillance, and enforced social distancing, as well as close monitoring of those infected and extensive quarantine and isolation measures. Authorities engaged in an unprecedented effort to cordon off Wuhan, the epicenter of the outbreak.

The measures adopted in China echoed and amplified approaches undertaken during recent outbreaks of SARS (2003) and MERS (2012). Although their intensity is unique, these measures are largely grounded in long-standing public health approaches to respiratory viruses. As COVID-19 has spread to other parts of the globe, other nations have adopted social distancing, quarantine, and isolation strategies. Here, we investigate how these measures have impacted COVID-19's time-varying reproduction number $R(t)$ at the national level and globally.

## Methods

We use a branching point process [1] to estimate a time-varying reproduction number $R(t)$ [2]. In particular, we model the intensity (rate) of infections as

(1) $\lambda(t) = \mu + \sum_{t > t_i} R(t_i) w(t - t_i)$,

where

(2) $p_{ij} = R(t_j) w(t_i - t_j) / \lambda(t_i)$

gives the probability of secondary infection i having been caused by primary infection j. Here, as in prior related works, the distribution of inter-event times $w(t_i - t_j)$ is modeled as Weibull. Because mitigation strategies (or the lack thereof) may influence the time distribution between primary and secondary infections, we estimate the Weibull parameters jointly with the model instead of specifying the parameters using reported incubation times. The baseline rate $\mu$ captures the rate of exogenous (imported) infections in a given region. The point process in Equation (1) can be viewed as an approximation to the common SIR model of infectious diseases during the initial phase of an epidemic when the total infections is small compared to the overall population size [3]. The model is estimated using a nonparametric expectation-maximization algorithm (details are included in the supplemental material) and a histogram estimator is used for $R(t)$.

Results

We analyze worldwide counts of COVID-19 deaths per day from January 22, 2020 to March 11, 2020 [4]. Because there is significant variability in testing levels across different countries and over time, we focus on mortality. In Figure 1 we plot estimated $R(t)$ along with 95% bootstrap confidence intervals. In China, we estimate $R$ was 1.50 (1.15, 1.79) on January 22, 2020 and as of March 11 is 0.43 (0.26, 0.34). In Italy, we estimate $R$ was 3.71 (2.00, 5.33) on February 21, 2020 and as of March 11 is 2.51 (2.24, 2.72). In Countries (aggregated) outside of China the estimated $R$ was 4.65 (3.17, 6.67) on February 21, 2020 and as of March 11 is 3.37 (3.14, 3.56).

Worldwide, the overall estimated $R$ was 2.17 (1.85, 2.49) on January 22, 2020. It decreased to 0.50 (0.39, 0.44) in mid February. By March 11, however, estimated $R$ had increased to 2.22 (2.09, 2.39). For comparison, recent research has estimated the average reproduction rate (across several studies) to be 3.28 (1.4, 6.5) [5]. A study conducted with data in China through mid February estimated that $R$ has been reduced from 2 to 1 [6].

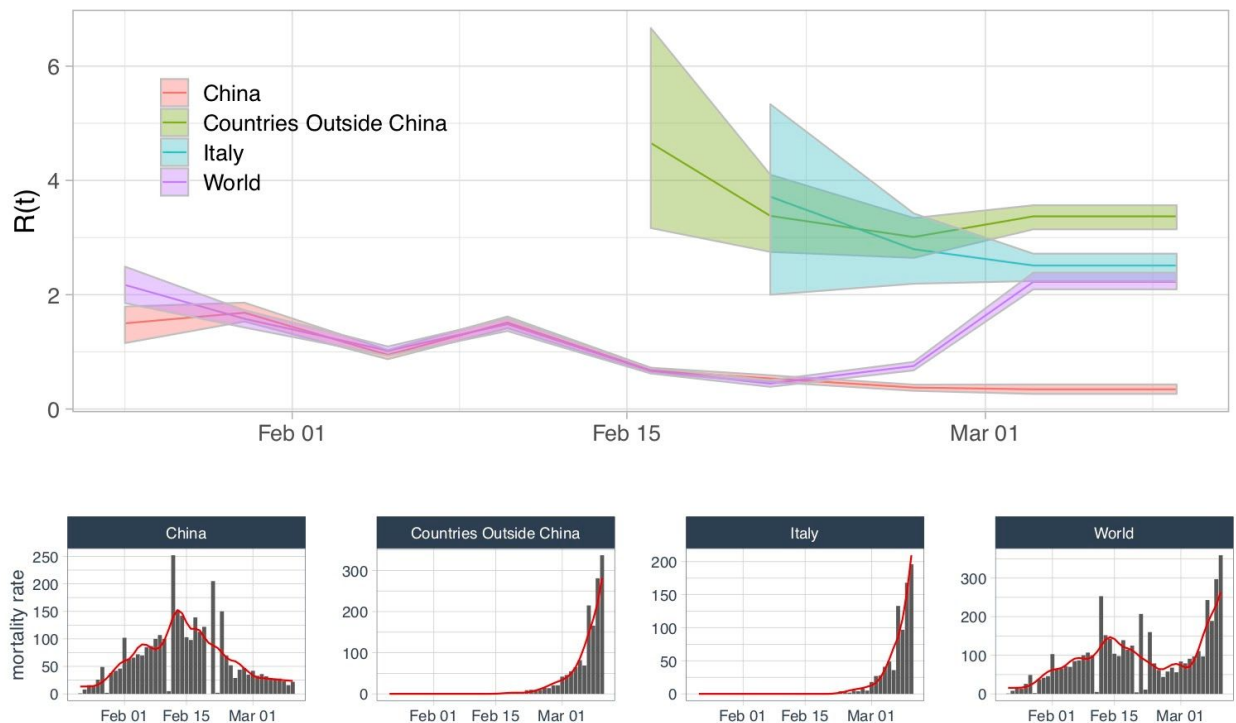

**Figure 1**. Top row: Estimated $R(t)$ over time and 95% confidence interval. Bottom row: Estimated point process rate of deaths along with death counts per day.

## Discussion

These results indicate that public health measures undertaken in China reduced the $R(t)$ of COVID-19 to below the self-sustaining level of 1 within China by the middle of February. They also suggest, however, the limitations of isolation, quarantine, and large-scale attempts to limit travel. While the world-wide $R(t)$ briefly dropped below 1 as China implemented extensive public health measures, the introduction of the virus to other nations swiftly led to an increasing world-wide average value of $R(t)$. In Italy, the nation hardest-hit following China, social distancing measures brought the local value of $R(t)$ down. Nonetheless, the value of $R(t)$ in Italy persisted at levels well above 1, allowing for ongoing transmission. By mid-March 2020, as COVID-19 spread in areas without extensive public health interventions in place, the world-wide value of $R(t)$ increased to a level similar to that of late January.

## Funding

This research was supported by NSF grants SCC-1737585, ATD-1737996 and ATD-1737925.

Supplemental Material

The intensity (rate) of events of the point process is defined as

(1) $\lambda(t) = \mu + \sum_{t > t_i} R(t_i) w(t - t_i)$.

The model can be represented as a branching process [s1] where the probability that infection i was caused by infection j is given by,

(2) $p_{ij} = R(t_j) w(t_i - t_j) / \lambda(t_i)$.

As in other related works [s2], we model the inter-time distribution $w(t_i - t_j)$ as Weibull with parameters $\alpha$ and $\beta$, which will be estimated from the data, and we model $R_0(t_j)$ as a piecewise constant function:

(3) $R(t) = \sum_{k=1}^{B} r_k 1\{t \in I_k\}$.

Here $I_k$ are intervals discretizing time, B is the number of such intervals, and $r_k$ is the estimated reproduction rate in interval k.

This representation facilitates an expectation-maximization (EM) algorithm for maximum likelihood inference [s3,s4,s5]. Given initial guesses for the model parameters $\mu$ $\alpha$ $\beta$ and $r_k$, the EM algorithm iteratively updates the parameters and branching probabilities by alternating between the

E-step update

$p_{ij} = R(t_j) w(t_i - t_j) / \lambda(t_i)$    $p_{ii} = \mu / \lambda(t_i)$

and M-step update

$w(t) \sim MLE_{Weibull}(\{t_i - t_j; p_{ij}\})$    $\mu = \sum_i p_{ii} / T$    $r_k = \sum_{i > j} p_{ij} 1\{t_j \in I_k\} / N_k$

where T is the total length of the observation period, $N_k$ is the total number of events in interval k, and the weibull is estimated via weighted MLE using the inter-event times as observations and branching probabilities as weights. To correct for the boundary where future secondary infections are not yet observed we estimate $r_B = r_{B-1}$.

We validate the model by generating synthetic data with parameters $\mu = 1$ $\alpha = 8$, $\beta = 4$ and $r_k$ given in Figure S1. We compare the model to the TD.R0 estimator provided in the R0 software package that estimates a time-varying R value. The method is similar to ours and can be viewed as an EM algorithm that is stopped after 1 iteration using initial guess $R(t) \equiv 1$ and using a histogram estimate with bin width equal to 1 day. We found the R0.TD method tends to produce noisy estimates, which is why we use the modified algorithm outlined here.

In order to assess goodness-of-fit we employ residual analysis of rescaled event times [s6]

$$\tau_i = \int_0^{t_i} \lambda(t)\, dt.$$

The rescaled times are distributed according to a unit rate Poisson process if the model is correctly specified. In Figure S2, we find that the estimated intensity provides a good fit for China and Italy mortality rates. The estimated intensity also provides a good fit for worldwide mortality rates with the exception of the final 2 days (March 10 and 11) where mortality counts were under-estimated by the model.

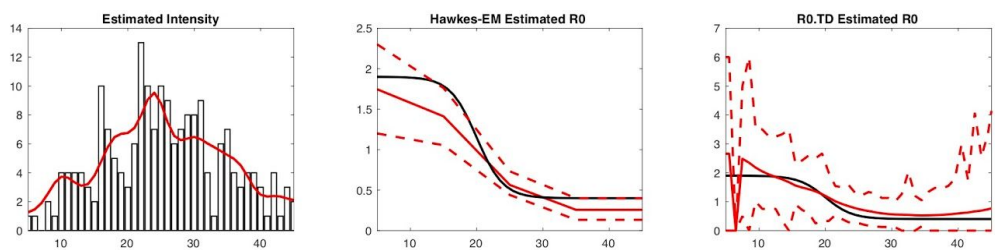

**Figure S1**: Model validation on synthetic data. Comparison of Hawkes-EM estimator of R(t) with the TD.R0 dynamic R(t) estimator in the ``R0'' R package. Left: histogram of event times of simulated Hawkes process (black) and estimated intensity (red) using Hawkes-EM algorithm (log-likelihood 127.0 compared to 122.3 for TD.R0). Center: R(t) estimate and 95% confidence interval for Hawkes-EM estimator (red) and ground truth R(t) (black). Right: R(t) estimate and 95% confidence interval for TD.R0 estimator (red) and ground truth R(t) (black).

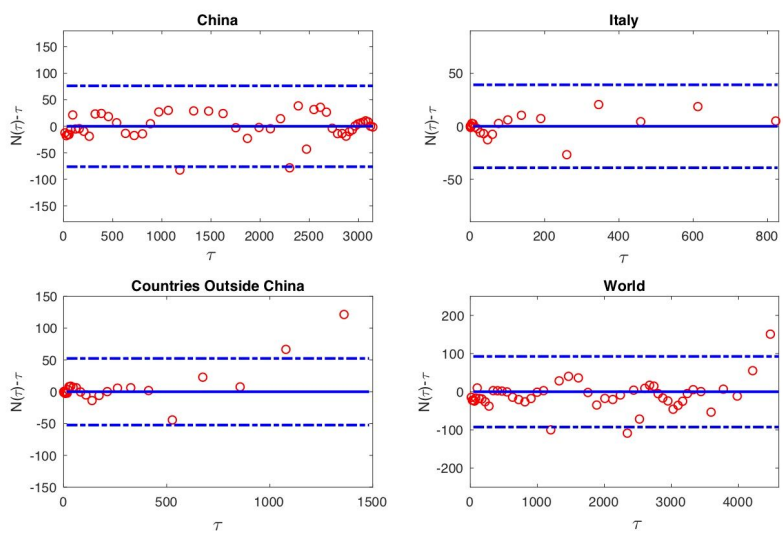

**Figure S2**: Normalized cumulative distribution of rescaled event times along with 95% error bounds of the Kolmogorov-Smirnov statistic.